\documentclass[12pt,a4paper,twoside]{article}
\usepackage{epsfig}
\usepackage{baltlat1}
\usepackage{wrapfig}

\begin{document}

\ \ \vspace{0.5mm}
\setcounter{page}{1} \vspace{8mm}
\titlehead{Baltic Astronomy, vol.12, XXX--XXX, 2003.}
\titleb{Physical Difference between the Short and Long GRBs}
\begin{authorl}
\authorb{L.G. Bal\'azs}{1},
\authorb{Z. Bagoly}{2},
\authorb{I. Horv\'ath}{3},
\authorb{A. M\'esz\'aros}{4} and
\authorb{P. M\'esz\'aros}{5}
\end{authorl}

\begin{addressl}
\addressb{1}{Konkoly Observatory, Budapest, Box 67, H-1525,
Hungary}

\addressb{2}{Laboratory for Information Technology, E\"{o}tv\"{o}s
University, P\'azm\'any P\'eter s\'et\'any 1/A, H-1518, Hungary}

\addressb{3}{Dept. of Physics, Bolyai Military University, Budapest, Box 12,
H-1456, Hungary}

\addressb{4}{Astronomical Institute of the Charles University,
180 00 Prague 8, V Hole\v sovi\v ck\'ach 2, Czech Republic}

\addressb{5}{Dept. of Astronomy \& Astrophysics, 
Pennsylvania State University, 525
Davey Lab, USA}
\end{addressl}
\submitb{Received October 15, 2003}
\begin{abstract}
We provided separate
bivariate log-normal distribution fits to the BATSE short and long
burst samples using the durations and fluences.
We show that these fits present an evidence for a
power-law dependence between the fluence and the duration, with a
statistically significant different index for the long and short
groups. We argue that the effect is probably real, and
the two subgroups are different physical phenomena. This may
provide a potentially useful constraint for models of long and
short bursts.
\end{abstract}
\begin{keywords}
gamma rays: bursts
\end{keywords}

\resthead{A Physical Difference ...}{L.~G. Bal\'azs et al.}
\sectionb{1}{INTRODUCTION}
The simplest grouping of gamma-ray bursts (GRBs)  divides bursts
into long ($T_{90} > 2s$) and short ($T_{90} < 2s$) duration
groups. Here we analyze (see also Bal\'azs et al. 2003)
the distribution of the observed fluences and
durations, and present arguments indicating that the intrinsic
durations and fluences are well represented by log-normal
distributions within these groups. We calculate the power-law
exponent  for the two types of bursts and the possible
implications for GRB models are briefly discussed. A statistically 
significant difference between the two subgroups will be found.

\sectionb{2}{ANALYSIS OF THE DURATION DISTRIBUTION}
Our GRB sample is selected from the current BATSE Gamma-Ray Burst
Catalog (Meegan et al. 2001). The bimodal distribution of the
observed $T_{90}$ can be well fitted by means of two Gaussian
distributions in the logarithmic durations (Horv\'ath 1998).  The
$T_{90}$ observed duration of  GRBs
 (which may be subject to cosmological time dilatation) relates
to $t_{90}$ the duration which would be measured by a comoving
observer by
\begin{equation}
T_{90} = t_{90}\, f(z),
\end{equation}
where z is the redshift, and $f(z)$ measures the time dilatation.
Taking the logarithms of both sides of equation one obtains the
logarithmic duration as a sum of two independent stochastic
variables. According to a theorem of  Cramer (Cram\'er 1937), if a
variable $\zeta$ which has a Gaussian distribution is given by the
sum of two independent variables, e.g. $\zeta= \xi + \eta$, then
both $\xi$ and $\eta$ have Gaussian distributions. The Gaussian
distribution of $\log T_{90}$ implies that the same type of
distribution exists for the variables $\log t_{90}$ and $\log
f(z)$. The distribution of $\log f(z)$ cannot be Gaussian which
means that the Gaussian nature of the distribution of $\log
T_{90}$ must be dominated by the distribution of $\log t_{90}$.
\sectionb{3}{DISTRIBUTION OF THE ENERGY FLUENCES}
The fluences are given in four different energy channels, $F_1$,
$F_2$, $F_3$, $F_4,$ whose energy bands correspond to [25,50] keV,
[50,100] keV, [100,300] keV and $>$ 300 keV. The total fluence is
defined as $F_{tot} = F_1 + F_2 + F_3 + F_4$. The observed total
fluence $F_{tot}$ can be expressed as 
\begin{equation}
F_{tot} = \frac{(1+z)}{4\pi d_l^2(z)} E_{tot} = c(z) E_{tot}.
\end{equation}
Here $E_{tot}$ is the total emitted energy of the GRB at the
source in ergs, the total fluence has dimension of erg/cm2, and
$d_l(z)$ is the luminosity distance corresponding to redshift $z$.
 Accepting
the hypothesis of a Gaussian distribution within the short and the
long group, one can apply again Cramer's theorem which leads to
the conclusion that either both the distribution of $\log c(z)$
and the distribution of $\log E_{tot}$ are Gaussian, or else the
variance of one of these quantities is negligible compared to the
other, which then must be mainly responsible for the Gaussian
behavior. Calculating  the variance of $\log c(z)$  for  the GRBs
with known redshifts  one obtains $\sigma_{\log c(z)}=0.43$. The
total observed variance is $\sigma_{\log F{tot}}=0.66$. Hence, the
variance of $c(z)$ gives roughly a $(0.43/0.66)^2 \approx 43\%$
contribution to the entire variance, so its significant fraction
is also intrinsic.
\sectionb{4}{RELATIONSHIP BETWEEN FLUENCE AND DURATION}
We observe only those bursts which fulfil some triggering criteria
and  the observed quantities are suffering from some type of bias
depending on the process of detection. The detection proceeds on
three time scales: on 64, 256 and 1024 ms resolution. We shown in
a previous paper (Bagoly et al. 1998) that the duration and the
peak intensity are independent stochastic variables:
\begin{equation}
\log F_{tot} = a_1 log T90 + a_2 log P + \varepsilon,
\end{equation}
where $a_1$ and $a_2$ are  constants and $\varepsilon$ is some
noise term. By fixing the $P$ peak intensity we get the
relationship between the fluence  and  duration. To ensure the
best time resolution we used $P_{64}$ in our calculations. In
order to fit the $[\log T_{90}, \log F_\mathrm{tot}]$ data pairs
with the superposition of two two-dimensional Gaussian bivariate
distributions we splitted the Catalog into subsamples with respect
to 64~ms peak fluxes.
The strata were obtained by taking 0.2 wide strips in the
logarithmic peak fluxes in the $0.0 < \log P_{64} < 1.0$ range.
With this choice we avoided the incompleteness among the faint
bursts and get enough objects for statistics within the strips
investigated. The calculated weighted mean of the slopes  resulted
in $a_1=0.81 \pm 0.06$ for the short bursts and  $a_1=1.11 \pm 0.03$
 for the long bursts. The difference is significant at the 99.998 \%
 level. Fig. 1 shows the results obtained for the faintest and
 brightest subsample of our calculations.
\begin{figure}
\vskip2mm
\centerline{{\psfig{figure=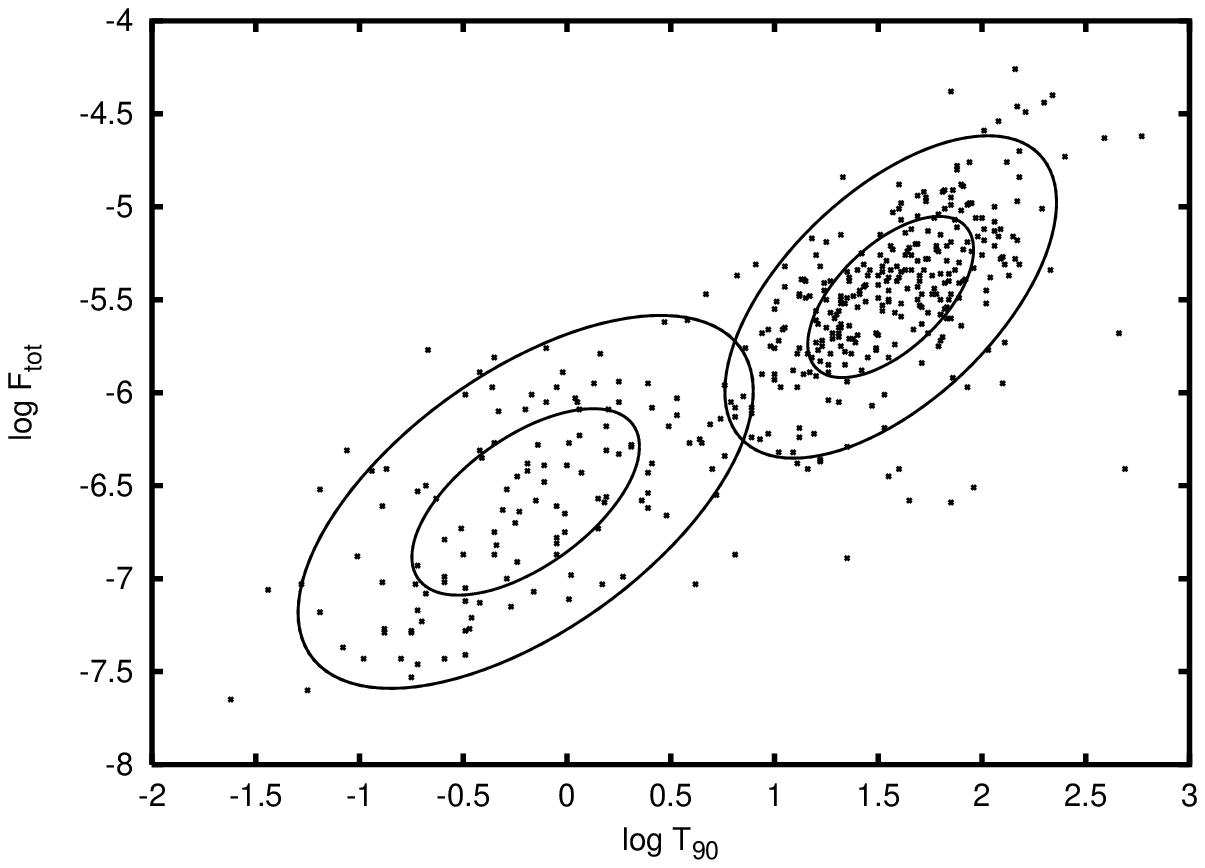,width=60truemm,angle=0,clip=}}
\hskip2mm
{\psfig{figure=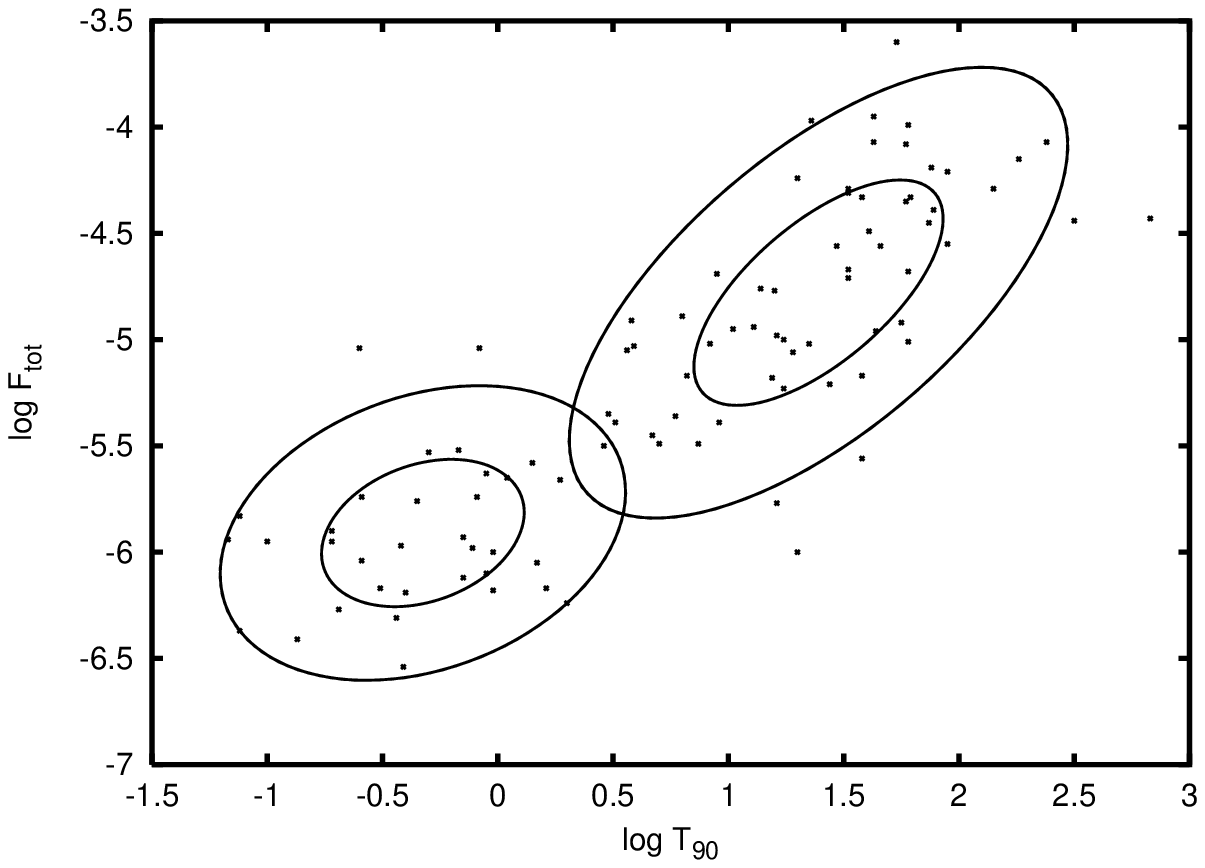,width=60truemm,angle=0,clip=}}}
\captionb{1}{The best $Maximum \, Likelihood$ fits of the two
log-Gaussian distributions for the faintest and brightest
subsample. The $1 \sigma$ and $2\sigma$ error ellipses of the two
components are indicated.} \vskip2mm
\end{figure}
\sectionb{5}{CONCLUSIONS}
We have presented evidence indicating that there is a power-law
relationship between the logarithmic fluences and the $T_{90}$
durations. An intriguing corollary of these results is that the
exponents in the power-law dependence between fluence and duration
differ significantly for the two groups of short ($T_{90} < 2s$)
and long ($T_{90} > 2s$) bursts. This also means that the same
power law relations hold between the total energy emitted
($E_{tot}$) and the intrinsic durations ($t_{90}$) of the two
groups. In summary, we have presented quantitative arguments that
there is a power law relation between the fluence and duration of
GRBs  and  this relation is significantly different for the two
groups of short and long bursts. This may indicate that two
different types of central engines are at work, or perhaps two
different types of progenitor systems are involved.
ACKNOWLEDGMENTS: This research was supported in part through OTKA
grants T024027 (L.G.B.), 
and T034549, NASA grant
NAG-9192 and NAG-9153, (P.M.), and Czech Research Grant J13/98:
113200004 (A.M.).
\References

 \ref Bal\'azs, L.G., Bagoly, Z., Horv\'ath, I., et al.
2003, A\&A, 410, 129 

\ref Bagoly, Z., M\'esz\'aros, A., Horv\'ath,
I., et al. 1998, ApJ, 498, 342 

\ref Cram\'er, H. 1937, Random
variables and probability distributions, Cambridge Tracts in
Mathematics and Mathematical Physics, No.36 (Cambridge University
Press, Cambridge) 

\ref Horv\'ath, I. 1998, ApJ, 508, 757

 \ref
Meegan, C.A., Pendleton, G.N, Briggs, M.S., et al. 2001, Current
BATSE Gamma-Ray Burst Catalog, \newline
http://gammaray.msfc.nasa.gov/batse/grb/catalog/current/
\end{document}